\documentclass[aps,prl,twocolumn,groupedaddress]{revtex4}
\usepackage{graphicx}
\input epsf
\usepackage{amsmath,amssymb}
\begin{document}

\title{Enhanced performance of MoS$_2$/SiO$_2$ field-effect transistors by hexamethyldisilazane (HMDS) encapsulation}
\author{Santu Prasad Jana}
\affiliation{Department of Physics, Indian Institute of Technology Kanpur, Kanpur 208016, India}
\author{Shivangi}
\affiliation{Department of Physics, Indian Institute of Technology Kanpur, Kanpur 208016, India}
\author{Suraina Gupta}
\affiliation{Department of Physics, Indian Institute of Technology Kanpur, Kanpur 208016, India}
\author{Anjan K. Gupta}
\affiliation{Department of Physics, Indian Institute of Technology Kanpur, Kanpur 208016, India}
\date{\today}

\begin{abstract}
Scalable methods for improving the performance and stability of a field-effect transistor (FET) based on two-dimensional materials are crucial for its real applications. A scalable method of encapsulating the exfoliated MoS$ _{2} $ on SiO$ _{2} $/Si substrate by hexamethyldisilazane (HMDS) is explored here for reducing the influence of interface traps and ambient contaminants. This leads to twenty-five times reduction in trap density, three times decrease in subthreshold swing, three times increase in the peak field-effect mobility and a drastic reduction in hysteresis. This performance remains nearly the same after several weeks of ambient exposure of the device. This is attributed to the superhydrophobic nature of HMDS and the SiO$_2$ surface hydrophobization by the formation of covalent bonds between the methyl groups of HMDS and silanol groups of SiO$_{2}$.
\end{abstract}
\maketitle
Semiconducting transition metal dichalcogenides (TMDs) \cite{TMDs} provide many advantages when integrated into field effect transistors (FETs) in their atomically thin form. Molybdenum disulfide (MoS$_2$) has appeared as the most favored and vastly researched semiconducting TMD in recent years due to its natural abundance as well as excellent environmental stability. Its mechanical flexibility, high transparency, thickness-dependent bandgap \cite{direct gap,direct gap1}, higher mobility than organic semiconductors and the electrostatic gate control make it a favored candidate for the next-generation nano-electronic devices. MoS$_2$'s applications have been demonstrated in many devices including transistor \cite{direct gap1,how good}, logic \cite{logic,logic1}, high-frequency \cite{high frequency}, circuit integration \cite{ic}, photodetector \cite{photodetectors} optoelectronic \cite{direct gap1,optical helicity}, light emitters \cite{light} and photovoltaic cells \cite{photo1}.

Nonetheless, the actual device performance of MoS$ _{2} $ FETs remain significantly below the intrinsic theoretical limit \cite{mobility,mobility1} with the actual devices exhibiting hysteresis and degradation \cite{hysteresis, scalling} with time, particularly, on exposure to ambient conditions. These traits can be attributed to either charge-traps at the MoS$ _{2} $/SiO$ _{2} $ interface \cite{interface traps} or the instability related to the facile adsorption of oxygen and water molecules \cite{BTI}. Consequently, in long-term, the subthreshold swing (SS) becomes large and the carrier mobility reduces. Passivation of the dielectric interface with MoS$ _{2} $ and protection of top surface by a capping layer have been tried by many research groups. These include different dielectric and encapsulation layers such as, Al$ _{2} $O$ _{3} $, HfO$ _{2} $, hBN, and PMMA \cite{Al2O3,Al2O31,Al2O32,hBN,HfO2,PMMA}. Despite significant improvement, owing to the lack of surface dangling bonds in MoS$ _{2} $, atomic layer deposition (ALD)-processed oxide capping layers exhibit non-uniform growth leading to partial coverage when the capping layer is ultra-thin. The most remarkable performance is achieved with exfoliated hBN encapsulation but this method is not scalable.

In this paper, we present a facile method, for efficient and scalable interface passivation and top protection, for few-layer MoS$_2$ with an air-stable and thin organic hexamethyldisilazane (HMDS) layer. The performance of an HMDS encapsulated device, operated in ambient conditions, is compared with the un-passivated device operated in vacuum. The HMDS encapsulated device exhibits twenty-five times less slow-trap density leading to negligible hysteresis, three times less sub-threshold swing and three times larger peak value of the field effect mobility as compared to the latter. An exposure to ambient air for 25 days leads to about 10\% mobility reduction with negligible hysteresis change. The effectiveness of this passivation and protection method is discussed in terms of HMDS properties.

Acetone and IPA cleaned highly p-doped Si wafers with 300 nm thermal SiO$_{2}$ are used as a substrate with the back gate. Single- or few-layer MoS$_{2}$ flakes are mechanically exfoliated from natural bulk crystal (from SPI) and transferred to the substrate. This uses the conventional dry transfer method \cite{XYZ} with the PDMS film (Gel film from Gel Pak) as a viscoelastic stamp and an XYZ-micromanipulator attached to an optical microscope. For interface passivation, acetone and isopropyl alcohol cleaned substrates were immersed in a 50:50 mixture of HMDS and acetone for 12-15 hrs followed by pure HMDS spin-coating at 2000 rpm for 45 s prior to the MoS$_{2}$ transfer. The source-drain contacts are made using mechanical masking with a 15 $\mu$m diameter tungsten wire after aligning the MoS$_{2}$ flake underneath it with the help of an optical microscope. This is followed by 50-nm-thick gold film deposition by thermal evaporation. By using mechanical masking, the organic lithography resist is avoided, which can leave residue on MoS$_{2}$. For the protection of MoS$_{2}$ on top, a second HMDS layer was spin coated after the gold contact deposition.
\begin{figure}[h]
	\centering
 	\includegraphics[width=3.4in]{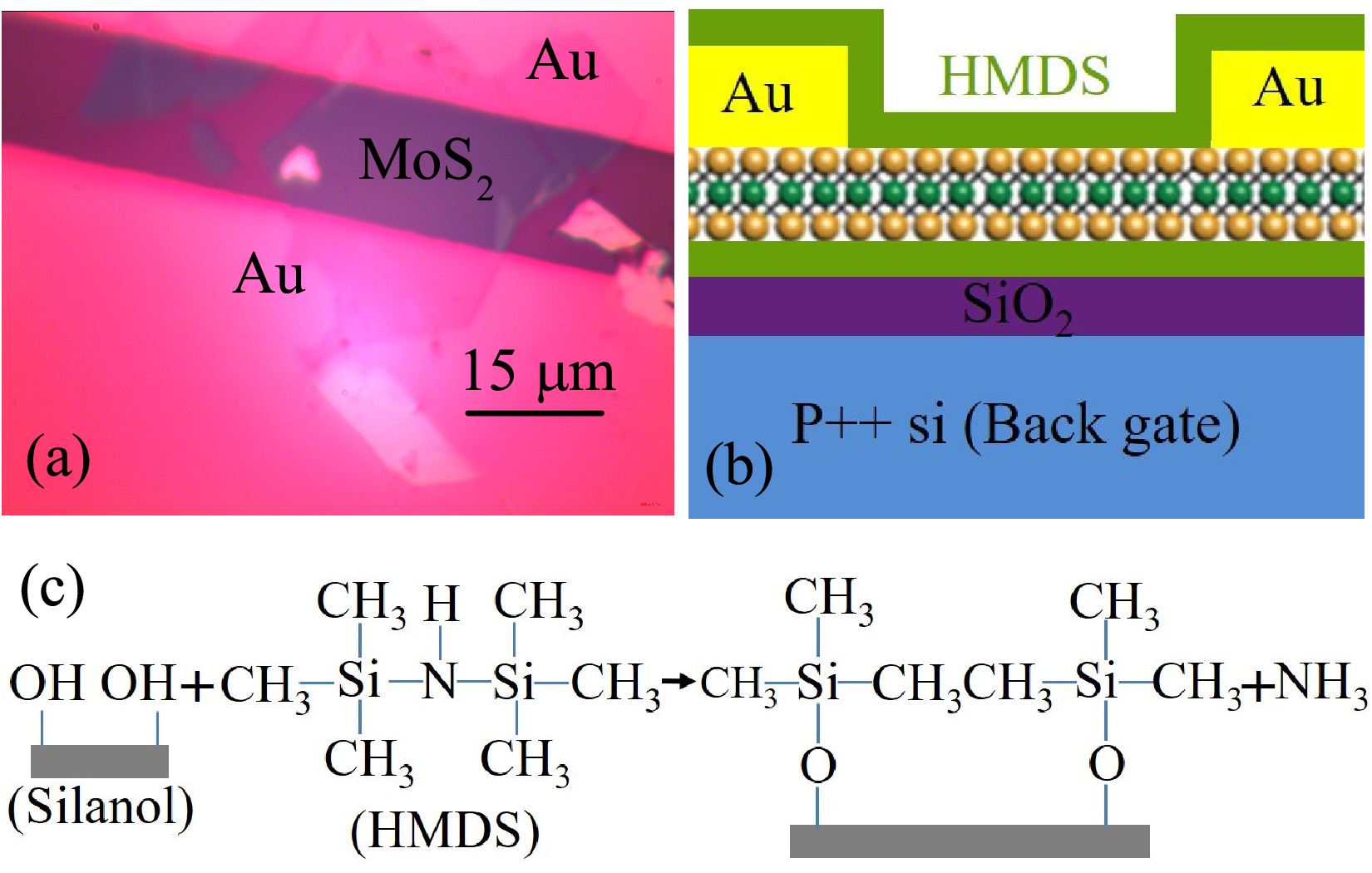}
	\caption{(a) shows an optical image of MoS$_2$ FET with HMDS encapsulation on SiO$_2$/Si substrate with gold contacts and (b) shows a schematic cross-section view of this device. (c) shows a diagrammatic illustration of the reaction between surface silanol groups and HMDS.}
	\label{fig:mos21}
\end{figure}

Figure \ref{fig:mos21}(a) shows the optical image of the few-layer MoS$_2$ FET with HMDS encapsulation and gold contacts. Since the HMDS layer is very thin and colorless, like other organosilicon compounds, the observed contrast and color of the MoS$_{2}$ flake on the ultra-thin HMDS layer do not differ much from that on bare SiO$_{2}$. Fig. \ref{fig:mos21}(b) shows the schematic of the device. Two probe conductance of both type, with and without HMDS encapsulation, devices were measured at room temperature in a homemade vacuum cryostat. A 10 k$\Omega$ series resistance was connected with the gate voltage $V_{\rm g}$ supply, which was controlled by a data acquisition card using a LabView program. The electrical measurements of without HMDS encapsulation device was done in vacuum in 10$^{-4}$ mbar range, whereas the HMDS encapsulated device was measured in the ambient conditions by keeping the cryostat cover open. The two probe transport used a drain-source voltage bias $V_{ds}$ controlled by a data acquisition card while the drain current $I_{\rm d}$ was measured through the voltage across a small bias-resistor, in series with the device, using a differential amplifier.

Thermally grown SiO$_2$ is an amorphous solid with dangling bonds and adsorbates acting as charge traps. When MoS$_2$ is placed on this surface, the interface trap states, which form within an accessible energy range of the channel's chemical potential, can change their occupancy and shield the gate electric field. Many different adsorbates can adhere to SiO$_2$. For instance, hydroxyl groups (-OH) pair with the surface-bound silicon's dangling bonds to form a layer of silanol (Si-OH) groups \cite{silica} that can act as electron traps. The silanol group can also do charge transfer with the MoS$_2$ via dipolar molecules such as water. Due to OH-termination, silanol is hydrophilic. Water molecules easily connect to the hydrogen of these silanol groups. Though some of the water molecules on the top surface of MoS$_2$ can be removed by vacuum annealing, a mono-layer or sub-mono-layer of hydrogen-bonded water cannot be extracted by pumping over long periods of time at room temperature \cite{silica, silica1}.

HMDS is a hydrophobic and air-stable organosilicon compound with the molecular formula [(CH$_3$)$_3$Si]$_2$NH, which is a derivative of ammonia with trimethylsilyl groups in place of two hydrogen atoms \cite{organosilicon,HMDS}. When HMDS is coated on the defective SiO$_2$ surface, the CH$_3$ in HMDS is covalently bonded with Si-OH \cite{HMDS1} and forms organosilyl by replacing the hydrogen of the free Si-OH group with organosilyl groups. The covalent reaction between the HMDS and surface silanol groups is shown in Fig.\ref{fig:mos21}(c). Thus, the SiO$_2$ surface becomes hydrophobic due to the HMDS molecules, leaving no hydroxyl group available to form hydrogen bonds with water molecules.

\begin{figure}[h]
\centering
\includegraphics[width=2.4in]{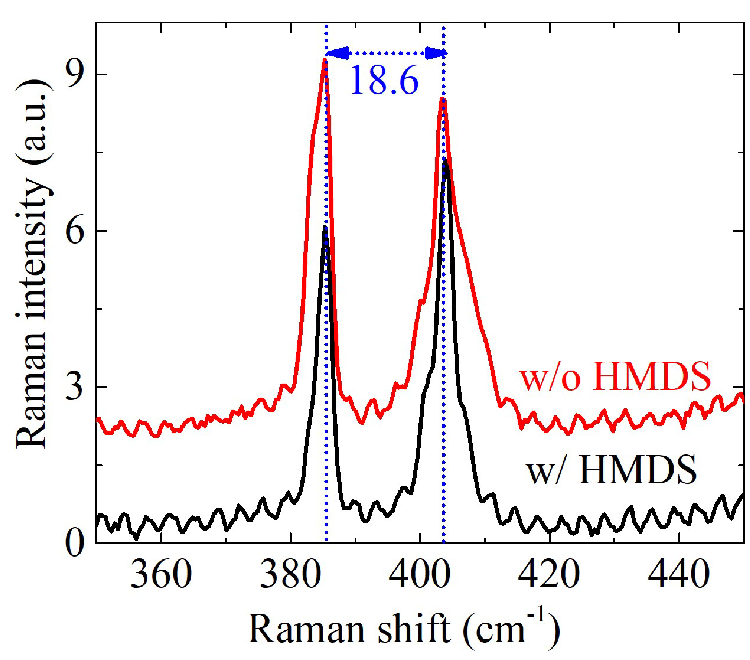}
\caption{Raman spectra of single-layer MoS$_2$ on SiO$_2$ with and without HMDS encapsulation showing sharpening of the Raman peaks resulting from the encapsulation.}
\label{fig:mos22}
\end{figure}

Figure \ref{fig:mos22} compares the Raman spectra of single-layer MoS$_2$ on SiO$_2$/Si with and without HMDS encapsulation. The two characteristic Raman peaks, namely $E^1_{\rm 2g} $ and $A_{\rm 1g} $ occur at 385.4 cm$^{-1} $ and at 404 cm$^{-1}$, respectively, for both cases, leading to a peak frequency separation of 18.6 cm$^{-1} $. This corresponds to single-layer MoS$_2$ as reported in literature \cite{anomalous lattice vibration, optical identification}.
The location of both the peaks in the HMDS-encapsulated MoS$_2$ layer remains same as that without HMDS within the experimental resolution. More significantly, in the HMDS-encapsulated MoS$_2$ layer, the full-width at half-maximum (FWHM) of the $E^1_{\rm 2g} $ and $A_{\rm 1g} $ peaks dropped from 4.15 cm$^{-1}$ to 2.08 cm$^{-1}$ and from 4.15 cm$^{-1} $ to 2.93 cm$^{-1} $, respectively. This suggests passivation of interface trap/defect states by HMDS as the interaction between the defect-bound excitons and phonons contributes to such line broadening.

\begin{figure}[h]
	\centering
 	\includegraphics[width=3.4in]{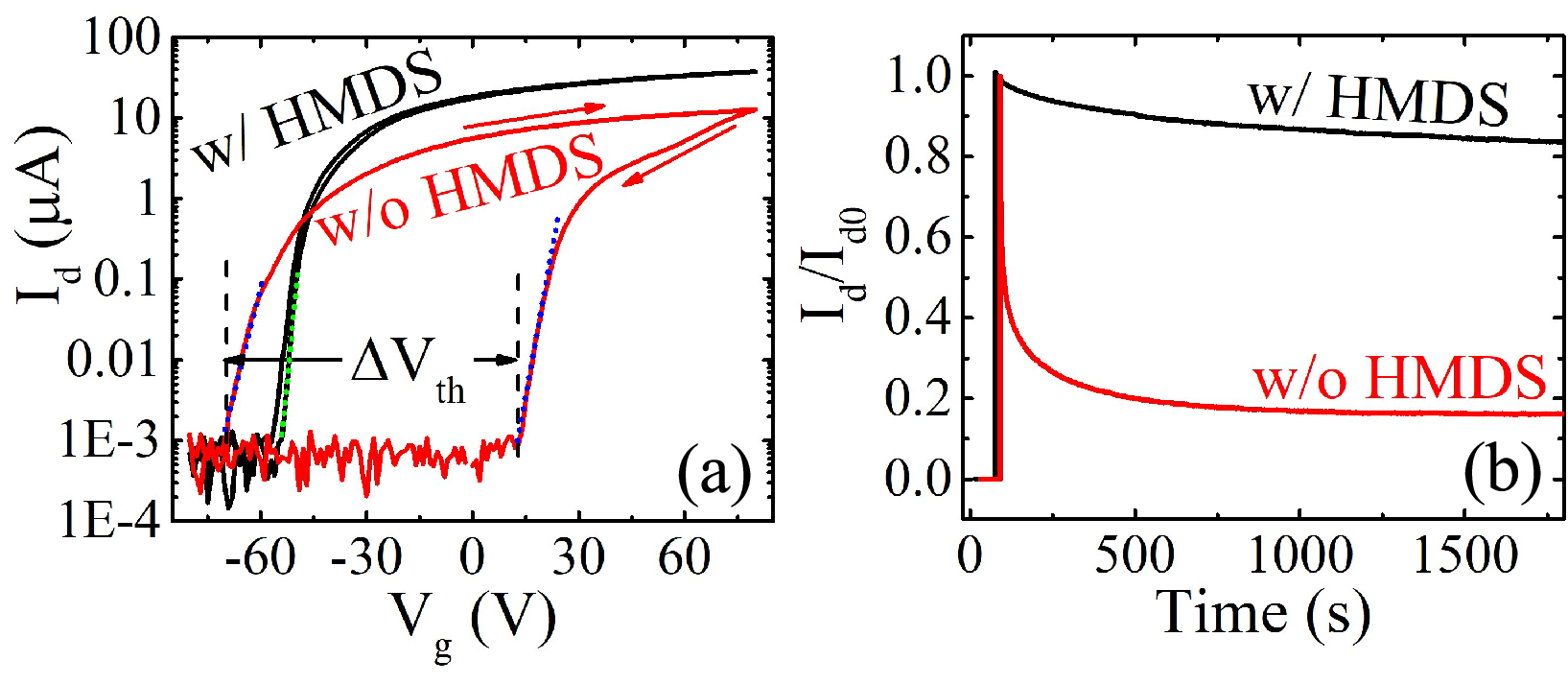}
	\caption{(a) $I_{\rm d}-V_{\rm g}$ transfer characteristics of MoS$_2$ FET at $V_{\rm ds}=1$ V with and without HMDS passivation. (b) The time-dependent $I_{\rm d}$ of MoS$_2$ FET with and without HMDS passivation when $V_{\rm g}$ is quickly altered from -80 to +80 V at t = 70 s and 80 s, respectively. The y-axis is normalized to show both the curves on the same plot.}
	\label{fig:mos23}
\end{figure}

Figure \ref{fig:mos23}(a) shows the $I_{\rm d}-V_{\rm g}$ curves of the few-layer MoS$_2$ devices with, and without, HMDS passivation. The increase in $I_{\rm d}$ with increasing $V_{\rm g}$ indicates that both devices exhibit the n-type conduction above certain threshold gate-voltage $V_{\rm th}$. The transfer characteristics without HMDS encapsulation, shown by the red curve in Fig. \ref{fig:mos23}(a), exhibits a large hysteresis, even in vacuum, with $V_{\rm th}$ that differ by $\Delta V_{\rm th}=82$ V during forward and reverse sweeps of $V_{\rm g}$. This large hysteresis is attributed to the charge-traps at the interface between MoS$_2$ and SiO$_2$ \cite{graphene-hyst,Bi-exponential,interface, blocking transition}. The areal density of the slow-traps responsible for this observed hysteresis is estimated using $n_{\rm str}=C_{\rm ox}\Delta V_{\rm th}/e$. Here, $e$ is the magnitude of electronic charge and $C_{\rm ox}= 12.1$ nFcm$^{-2}$ is the per-unit-area capacitance of SiO$_2$ layer. This leads to $n_{\rm str}= 6.2 \times 10^{12}$ cm$^{\rm-2}$.

The sub-threshold swing, defined as ${\rm SS}=1/(d \log I_{\rm d}/d V_{\rm g})$ just above $V_{\rm th}$, is calculated to be 2.9 V/dec for the device without HMDS. This is from the inverse of the slope of the blue dashed line in Fig.\ref{fig:mos23}(a) for the backward gate sweep. This SS can be used to estimate the density of states (DOS) $g_{\rm ftr}$ of the fast-traps using ${\rm SS}=k_{\rm b}T\ln10(1+\gamma_{\rm ftr})$. Here, $\gamma_{\rm ftr}=e^2g_{\rm ftr}/C_{\rm ox}$ is the ratio of the traps' quantum capacitance to the gate-oxide capacitance. This yields $g_{\rm ftr}= 3.5\times 10^{12}$ eV$^{-1}$cm$^{-2}$ for the device without HMDS.

The solid black line in Fig.\ref{fig:mos23}(a) is the transfer characteristic of the HMDS encapsulated device measured in ambient conditions. It is noteworthy that even in ambient conditions and over the same $V_{\rm g}$-sweep range and rate, the hysteresis is negligible. The $\Delta V_{\rm th}$ is reduced from 82 V to a mere 3 V, corresponding to a reduction in slow traps' density to 2.2 $\times$ 10$^{11}$ cm$^{\rm-2}$ i.e. more than 25 factor reduction as compared to the device without HMDS encapsulation. Further, the SS decreases from 2.9 V/dec to 0.9 V/dec, amounting to a fast trap DOS of 1.1 $\times$ 10$^{12} eV^{-1}cm^{-2}$, i.e. more than three factor reduction. Thus, the HMDS encapsulation significantly helps in passivating the interface traps as well as gate bias stress due to ambient species. Although the extracted SS value is far from the temperature-limited lowest value, i.e. 60 mV/dec with no traps, it is one of the lowest values for the 300 nm SiO$_2$ gate. Note that for a fixed trap DOS, the SS value can also be reduced by increasing the gate capacitance.

Figure \ref{fig:mos23}(b) shows $I_{\rm d}$ plots, scaled with their respective peak values, for the two devices as a function of time when $V_{\rm g}$ undergoes a step change from -80 to +80 V. $V_{\rm g}$ was held at -80 V for 1 hour to achieve equilibrium between the chemical potential of the trap states and that of the channel. At this step rise in $V_{\rm g}$ from -80V to 80V, the $I_{\rm d}$ has step rise, in both the devices, due to a sudden rise of carrier density in the channel. This is followed by an abrupt decrease in $I_{\rm d}$, especially in device without HMDS encapsulation, followed by a slow decrease in $I_{\rm d}$ due to the charging of the interface traps. The $I_{\rm d}$ of the unpassivated device decreases by almost 85\% of its value immediately after the voltage step, whereas the passivated device only shows a 16\% reduction, over 30 min time. This indicates a large areal density of traps, which capture the electrons from the channel, in the unpassivated device, leading to significant shielding of the back-gate electric field, as compared to the HMDS encapsulated device.

\begin{figure}[h]
	\centering
 	\includegraphics[width=3.4in]{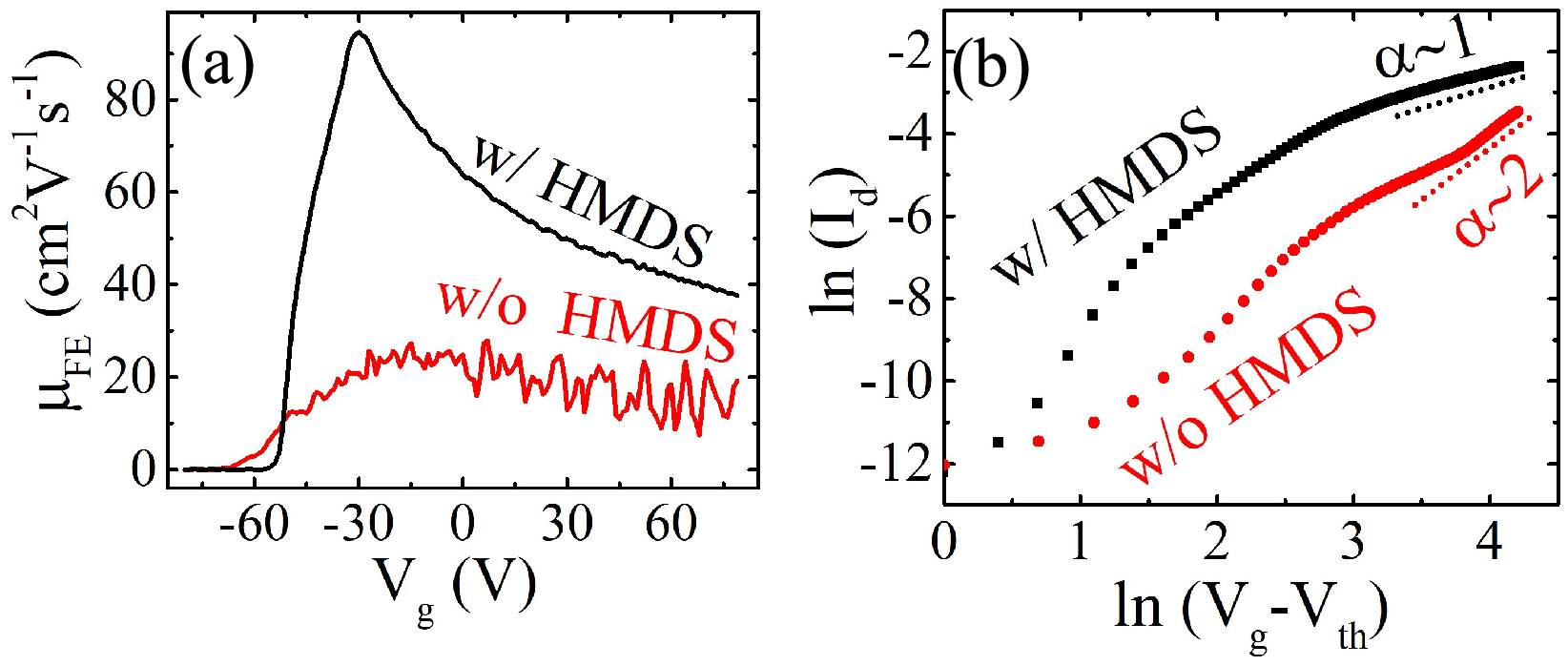}
	\caption{(a)The field-effect mobility
MoS$_2$ FET extracted from $I_{\rm d}-V_{\rm g}$ curves with and without HMDS passivation. (b) Variation of $I_{\rm d}$ with $V_{\rm g}-V_{\rm th}$ for MoS$_2$ FET with and without HMDS passivation.}
	\label{fig:mos24}
\end{figure}

Figure \ref{fig:mos24} shows the field-effect mobility $\mu_{\rm FE}$ of the MoS$_2$ FETs as a function of $V_{\rm g}$ with and without HMDS encapsulation with a marked difference between the two. The $\mu_{\rm FE}$ is extracted from the transfer characteristics using the definition $\mu_{\rm FE}=\left(\frac{L}{WC_{\rm ox}V_{\rm d}}\right)\left(\frac{dI_{\rm d}}{dV_{\rm g}}\right)$. Here, $W$ and $L$ are the channel width and length, respectively. Note that the maximum field-effect mobility of the HMDS encapsulated device is about 3.5 times of that of the unpassivated device. Note also the abrupt jumps in the $\mu_{\rm FE}$ of the device without HMDS, presumably due to a large number of traps changing their charge state at certain $V_{\rm g}$ values, while $\mu_{\rm FE}$ for device with HMDS is rather smooth.

The above $\mu_{\rm FE}$ ignores the effect of the quantum capacitance of the channel $C_{\rm ch}$ and that of the interface traps $C_{\rm tr}$. If one incorporates these, the change in channel carrier density, in response to $\Delta V_{\rm g}$ change in the gate voltage, is given by $\Delta n=\left(\frac{C_{\rm ox}\Delta V_{\rm g}}{e}\right)\left(\frac{C_{\rm ch}}{C_{\rm ch}+C_{\rm ox}+C_{\rm tr}}\right)$. This will lead to a more appropriate mobility expression $\mu=e^{-1}(\frac {dG}{dn})$ with $G$ as the channel conductivity, which will differ from $\mu_{\rm FE}$ by $\left(\frac{C_{\rm ch}}{C_{\rm ch}+C_{\rm ox}+C_{\rm tr}}\right)$ factor. Thus $\mu_{\rm FE}$ will correspond to the actual mobility in the limit $C_{\rm ch}\gg C_{\rm tr}$, $C_{\rm ox}$. This can be expected to be the case close to the degenerate limit. Thus $\mu_{\rm FE}$ can be assumed to be close to the actual mobility for large $(V_{\rm g}-V_{\rm th})$ values.

The overall dependence of mobility on the carrier density also reflects the nature of the carrier scattering. The scattering can occur from the Coulomb potential of the interface traps, phonons and other excitations, as well as from the intrinsic defects. For a Coulomb scatterer one can write $G\propto n^{\rm \alpha}$. Here, $1 \leq \alpha \leq 2$ is a parameter that depends on the screening of the Coulomb scatterer. For bare impurity Coulomb scattering, $\alpha=2$ and for screened Coulomb impurity scattering $\alpha=1$ \cite{impurity scattering,impurity scattering1}.

Close to the degenerate limit, $n$ can be estimated as $n=C_{ox}(V_{g}-V_{th})/e$.
With $G \propto I_{\rm d}$, the slope of the $\ln(I_{\rm d})$ - Vs - $\ln(V_{\rm g}-V_{\rm th})$ plot for large $V_{\rm g}$ can be used as an estimate of $\alpha$. From Fig. \ref{fig:mos24}, $\alpha \sim1.9$, \emph{i.e.} nearly 2, for the device without HMDS encapsulation, while for the HMDS encapsulated device, $\alpha \sim1.1$, \emph{i.e.} nearly 1. This analysis suggests that the HMDS encapsulation, or just the reduction in traps' density due to HMDS, reduces the bare Coulomb impurity scattering, presumably due to the reduction in interface traps, as compared to the screened impurity scattering.

A 2D semiconductor FET with its exposed channel to ambient air can accumulate more traps with time degrading device performance and causing threshold voltage instability. Besides these traps, there are also water and oxygen molecules in the environment that are absorbed depending on $V_{\rm g}$ value and cause the gate bias stress and lead to increased hysteresis. The latter is clearly insignificant in HMDS encapsulated device operated in air, as seen above, ruling out the gate bias stress.
\begin{figure}[h]
	\centering
 	\includegraphics[width=2.7in]{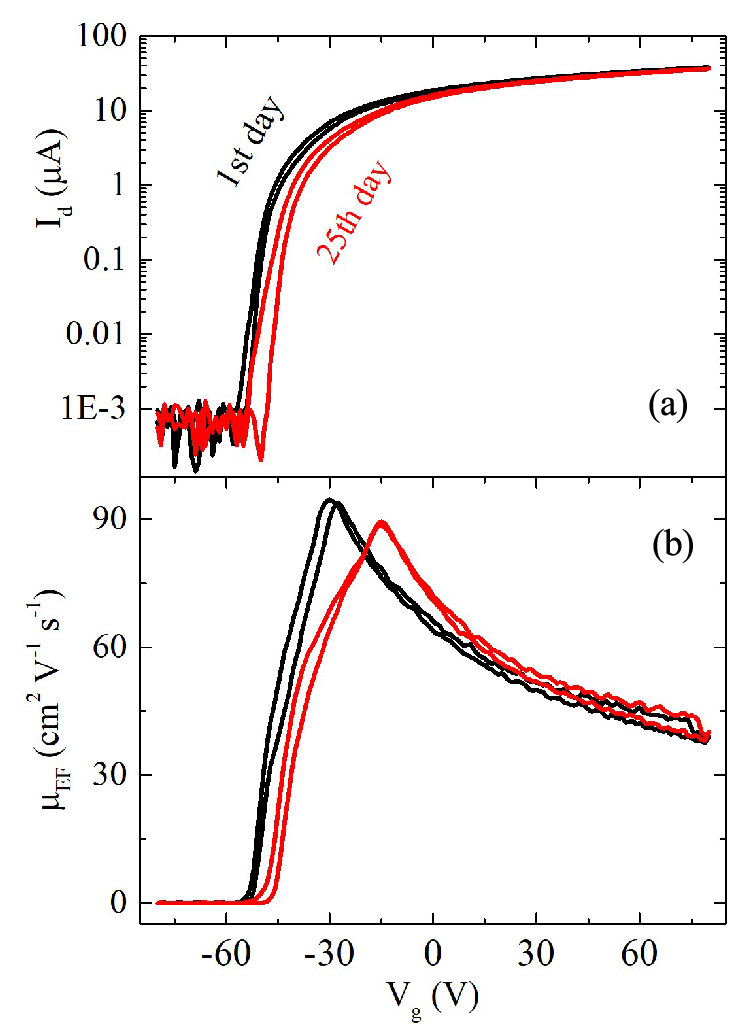}
	\caption{(a) Transfer characteristics of HMDS-passivated MoS$_2$ FET measured on 1st and 25th day. (b) The field-effect mobility extracted from $I_{\rm d}-V_{\rm g}$ curves of (a).}
	\label{fig:mos25}
\end{figure}
Further, the transfer characteristics of the HMDS encapsulated device, measured on the first day and after 25 days of keeping the device in air, are shown in Fig.\ref{fig:mos25}(a). There is negligible variation in the hysteresis window, a slight decrease in the current value, and a slight increase in the threshold voltage. The extracted $\mu_{\rm FE}$ in Fig.\ref{fig:mos25}(b), indicates a mere 10\% reduction after 25 days. This indicates that the top HMDS also protects MoS$_2$ channel quite well from the environmental oxygen and water molecules.

In conclusion, there is a substantial decrease in interface trap density and gate bias stress leading to an improved and lasting performance of MoS$_2$/SiO$_2$ FETs by HMDS encapsulation through interface passivation and top protection. A 25 factor reduction in slow traps' density, three times reduction in subthreshold swing and a significant improvement in the field-effect mobility is found in MoS$_2$ FET after HMDS encapsulation. The encapsulation method, based on spin coating of the chemically inert HMDS, is simple, scalable and can be extended to other 2D materials.


\end{document}